
%
%

%
%
%
%

\def\Serif{cmr}
\def\SerifBold{cmbx}
\def\SerifItalics{cmti}
\def\SerifSlanted{cmsl}
\def\SerifBoldItalics{cmbxti}
\def\SansSerif{cmss}
\def\SansSerifBold{cmssbx}
\def\SansSerifItalics{cmssi}
\def\SansSerifSlanted{cmssi}
\def\Math{cmmi}
\def\Symbols{cmsy}
\def\MathBold{cmmib}
\def\MoreSymbols{cmex}
\def\Typewriter{cmtt}
\def\Gothic{eufm}
\def\Double{msbm}
\def\Relazioni{msam}

= 			\Serif10 			at 5pt
= 		\SerifBold10 		at 5pt
= 	\SerifItalics10 	at 5pt
=		\SerifSlanted10 	at 5pt
=	\SerifBoldItalics10	at 5pt
= 		\SansSerif10 		at 5pt
=	\SansSerifBold10	at 5pt
=	\SansSerifItalics10	at 5pt
=	\SansSerifSlanted10	at 5pt
=				\Math10				at 5pt
=			\MathBold10			at 5pt
=			\Symbols10			at 5pt
=		\MoreSymbols10		at 5pt
=		\Typewriter10		at 5pt
=			\Gothic10			at 5pt
=			\Double10			at 5pt

= 			\Serif10 			at 7pt
= 		\SerifBold10 		at 7pt
= 	\SerifItalics10 	at 7pt
=	\SerifSlanted10 	at 7pt
=\SerifBoldItalics10	at 7pt
= 		\SansSerif10 		at 7pt
= 	\SansSerifBold10 	at 7pt
=\SansSerifItalics10	at 7pt
=\SansSerifSlanted10	at 7pt
=			\Math10				at 7pt
=		\MathBold10			at 7pt
=			\Symbols10			at 7pt
=		\MoreSymbols10		at 7pt
=		\Typewriter10		at 7pt
=			\Gothic10			at 7pt
=			\Double10			at 7pt

= 			\Serif10 			at 8pt
= 		\SerifBold10 		at 8pt
= 	\SerifItalics10 	at 8pt
=	\SerifSlanted10 	at 8pt
=\SerifBoldItalics10	at 8pt
= 		\SansSerif10 		at 8pt
= 	\SansSerifBold10 	at 8pt
=\SansSerifItalics10 at 8pt
=\SansSerifSlanted10 at 8pt
=			\Math10				at 8pt
=		\MathBold10			at 8pt
=			\Symbols10			at 8pt
=		\MoreSymbols10		at 8pt
=		\Typewriter10		at 8pt
=			\Gothic10			at 8pt
=			\Double10			at 8pt

= 			\Serif10 			at 10pt
= 		\SerifBold10 		at 10pt
= 		\SerifItalics10 	at 10pt
=		\SerifSlanted10 	at 10pt
=	\SerifBoldItalics10	at 10pt
= 		\SansSerif10 		at 10pt
= 	\SansSerifBold10 	at 10pt
= 	\SansSerifItalics10 at 10pt
= 	\SansSerifSlanted10 at 10pt
=				\Math10				at 10pt
=			\MathBold10			at 10pt
=			\Symbols10			at 10pt
=		\MoreSymbols10		at 10pt
=		\Typewriter10		at 10pt
=			\Gothic10			at 10pt
=			\Double10			at 10pt
=			\Relazioni10			at 10pt

= 				\Serif10 			at 12pt
= 			\SerifBold10 		at 12pt
= 		\SerifItalics10 	at 12pt
=		\SerifSlanted10 	at 12pt
=	\SerifBoldItalics10	at 12pt
= 			\SansSerif10 		at 12pt
= 		\SansSerifBold10 	at 12pt
= 	\SansSerifItalics10 at 12pt
= 	\SansSerifSlanted10 at 12pt
=				\Math10				at 12pt
=			\MathBold10			at 12pt
=			\Symbols10			at 12pt
=		\MoreSymbols10		at 12pt
=			\Typewriter10		at 12pt
=				\Gothic10			at 12pt
=				\Double10			at 12pt

= 			\Serif10 			at 14pt
= 		\SerifBold10 		at 14pt
= 	\SerifItalics10 	at 14pt
=		\SerifSlanted10 	at 14pt
=	\SerifBoldItalics10	at 14pt
= 		\SansSerif10 		at 14pt
= 	\SansSerifBold10 	at 14pt
= \SansSerifSlanted10 at 14pt
= \SansSerifItalics10 at 14pt
=				\Math10				at 14pt
=			\MathBold10			at 14pt
=			\Symbols10			at 14pt
=		\MoreSymbols10		at 14pt
=		\Typewriter10		at 14pt
=			\Gothic10			at 14pt
=			\Double10			at 14pt

\def\NormalStyle{\parindent=5pt\parskip=3pt\normalbaselineskip=14pt%
\def\nt{\tenSerif}%
\def\rm{\fam0\tenSerif}%
\textfont0=\tenSerif\scriptfont0=\sevenSerif\scriptscriptfont0=\fiveSerif
\textfont1=\tenMath\scriptfont1=\sevenMath\scriptscriptfont1=\fiveMath
\textfont2=\tenSymbols\scriptfont2=\sevenSymbols\scriptscriptfont2=\fiveSymbols
\textfont3=\tenMoreSymbols\scriptfont3=\sevenMoreSymbols\scriptscriptfont3=\fiveMoreSymbols
\textfont\itfam=\tenSerifItalics\def\it{\fam\itfam\tenSerifItalics}%
\textfont\slfam=\tenSerifSlanted\def\sl{\fam\slfam\tenSerifSlanted}%
\textfont\ttfam=\tenTypewriter\def\tt{\fam\ttfam\tenTypewriter}%
\textfont\bffam=\tenSerifBold%
\def\bf{\fam\bffam\tenSerifBold}\scriptfont\bffam=\sevenSerifBold\scriptscriptfont\bffam=\fiveSerifBold%
\def\cal{\tenSymbols}%
\def\greekbold{\tenMathBold}%
\def\gothic{\tenGothic}%
\def\Bbb{\tenDouble}%
\def\LieFont{\tenSerifItalics}%
\nt\normalbaselines\baselineskip=14pt%
}

\def\TitleStyle{\parindent=0pt\parskip=0pt\normalbaselineskip=15pt%
\def\nt{\fourteenSansSerifBold}%
\def\rm{\fam0\fourteenSansSerifBold}%
\textfont0=\fourteenSansSerifBold\scriptfont0=\tenSansSerifBold\scriptscriptfont0=\eightSansSerifBold
\textfont1=\fourteenMath\scriptfont1=\tenMath\scriptscriptfont1=\eightMath
\textfont2=\fourteenSymbols\scriptfont2=\tenSymbols\scriptscriptfont2=\eightSymbols
\textfont3=\fourteenMoreSymbols\scriptfont3=\tenMoreSymbols\scriptscriptfont3=\eightMoreSymbols
\textfont\itfam=\fourteenSansSerifItalics\def\it{\fam\itfam\fourteenSansSerifItalics}%
\textfont\slfam=\fourteenSansSerifSlanted\def\sl{\fam\slfam\fourteenSerifSansSlanted}%
\textfont\ttfam=\fourteenTypewriter\def\tt{\fam\ttfam\fourteenTypewriter}%
\textfont\bffam=\fourteenSansSerif%
\def\bf{\fam\bffam\fourteenSansSerif}\scriptfont\bffam=\tenSansSerif\scriptscriptfont\bffam=\eightSansSerif%
\def\cal{\fourteenSymbols}%
\def\greekbold{\fourteenMathBold}%
\def\gothic{\fourteenGothic}%
\def\Bbb{\fourteenDouble}%
\def\LieFont{\fourteenSerifItalics}%
\nt\normalbaselines\baselineskip=15pt%
}

\def\PartStyle{\parindent=0pt\parskip=0pt\normalbaselineskip=15pt%
\def\nt{\fourteenSansSerifBold}%
\def\rm{\fam0\fourteenSansSerifBold}%
\textfont0=\fourteenSansSerifBold\scriptfont0=\tenSansSerifBold\scriptscriptfont0=\eightSansSerifBold
\textfont1=\fourteenMath\scriptfont1=\tenMath\scriptscriptfont1=\eightMath
\textfont2=\fourteenSymbols\scriptfont2=\tenSymbols\scriptscriptfont2=\eightSymbols
\textfont3=\fourteenMoreSymbols\scriptfont3=\tenMoreSymbols\scriptscriptfont3=\eightMoreSymbols
\textfont\itfam=\fourteenSansSerifItalics\def\it{\fam\itfam\fourteenSansSerifItalics}%
\textfont\slfam=\fourteenSansSerifSlanted\def\sl{\fam\slfam\fourteenSerifSansSlanted}%
\textfont\ttfam=\fourteenTypewriter\def\tt{\fam\ttfam\fourteenTypewriter}%
\textfont\bffam=\fourteenSansSerif%
\def\bf{\fam\bffam\fourteenSansSerif}\scriptfont\bffam=\tenSansSerif\scriptscriptfont\bffam=\eightSansSerif%
\def\cal{\fourteenSymbols}%
\def\greekbold{\fourteenMathBold}%
\def\gothic{\fourteenGothic}%
\def\Bbb{\fourteenDouble}%
\def\LieFont{\fourteenSerifItalics}%
\nt\normalbaselines\baselineskip=15pt%
}

\def\ChapterStyle{\parindent=0pt\parskip=0pt\normalbaselineskip=15pt%
\def\nt{\fourteenSansSerifBold}%
\def\rm{\fam0\fourteenSansSerifBold}%
\textfont0=\fourteenSansSerifBold\scriptfont0=\tenSansSerifBold\scriptscriptfont0=\eightSansSerifBold
\textfont1=\fourteenMath\scriptfont1=\tenMath\scriptscriptfont1=\eightMath
\textfont2=\fourteenSymbols\scriptfont2=\tenSymbols\scriptscriptfont2=\eightSymbols
\textfont3=\fourteenMoreSymbols\scriptfont3=\tenMoreSymbols\scriptscriptfont3=\eightMoreSymbols
\textfont\itfam=\fourteenSansSerifItalics\def\it{\fam\itfam\fourteenSansSerifItalics}%
\textfont\slfam=\fourteenSansSerifSlanted\def\sl{\fam\slfam\fourteenSerifSansSlanted}%
\textfont\ttfam=\fourteenTypewriter\def\tt{\fam\ttfam\fourteenTypewriter}%
\textfont\bffam=\fourteenSansSerif%
\def\bf{\fam\bffam\fourteenSansSerif}\scriptfont\bffam=\tenSansSerif\scriptscriptfont\bffam=\eightSansSerif%
\def\cal{\fourteenSymbols}%
\def\greekbold{\fourteenMathBold}%
\def\gothic{\fourteenGothic}%
\def\Bbb{\fourteenDouble}%
\def\LieFont{\fourteenSerifItalics}%
\nt\normalbaselines\baselineskip=15pt%
}

\def\SectionStyle{\parindent=0pt\parskip=0pt\normalbaselineskip=13pt%
\def\nt{\twelveSansSerifBold}%
\def\rm{\fam0\twelveSansSerifBold}%
\textfont0=\twelveSansSerifBold\scriptfont0=\eightSansSerifBold\scriptscriptfont0=\eightSansSerifBold
\textfont1=\twelveMath\scriptfont1=\eightMath\scriptscriptfont1=\eightMath
\textfont2=\twelveSymbols\scriptfont2=\eightSymbols\scriptscriptfont2=\eightSymbols
\textfont3=\twelveMoreSymbols\scriptfont3=\eightMoreSymbols\scriptscriptfont3=\eightMoreSymbols
\textfont\itfam=\twelveSansSerifItalics\def\it{\fam\itfam\twelveSansSerifItalics}%
\textfont\slfam=\twelveSansSerifSlanted\def\sl{\fam\slfam\twelveSerifSansSlanted}%
\textfont\ttfam=\twelveTypewriter\def\tt{\fam\ttfam\twelveTypewriter}%
\textfont\bffam=\twelveSansSerif%
\def\bf{\fam\bffam\twelveSansSerif}\scriptfont\bffam=\eightSansSerif\scriptscriptfont\bffam=\eightSansSerif%
\def\cal{\twelveSymbols}%
\def\bg{\twelveMathBold}%
\def\gothic{\twelveGothic}%
\def\Bbb{\twelveDouble}%
\def\LieFont{\twelveSerifItalics}%
\nt\normalbaselines\baselineskip=13pt%
}

\def\SubSectionStyle{\parindent=0pt\parskip=0pt\normalbaselineskip=13pt%
\def\nt{\twelveSansSerifItalics}%
\def\rm{\fam0\twelveSansSerifItalics}%
\textfont0=\twelveSansSerifItalics\scriptfont0=\eightSansSerifItalics\scriptscriptfont0=\eightSansSerifItalics%
\textfont1=\twelveMath\scriptfont1=\eightMath\scriptscriptfont1=\eightMath%
\textfont2=\twelveSymbols\scriptfont2=\eightSymbols\scriptscriptfont2=\eightSymbols%
\textfont3=\twelveMoreSymbols\scriptfont3=\eightMoreSymbols\scriptscriptfont3=\eightMoreSymbols%
\textfont\itfam=\twelveSansSerif\def\it{\fam\itfam\twelveSansSerif}%
\textfont\slfam=\twelveSansSerifSlanted\def\sl{\fam\slfam\twelveSerifSansSlanted}%
\textfont\ttfam=\twelveTypewriter\def\tt{\fam\ttfam\twelveTypewriter}%
\textfont\bffam=\twelveSansSerifBold%
\def\bf{\fam\bffam\twelveSansSerifBold}\scriptfont\bffam=\eightSansSerifBold\scriptscriptfont\bffam=\eightSansSerifBold%
\def\cal{\twelveSymbols}%
\def\greekbold{\twelveMathBold}%
\def\gothic{\twelveGothic}%
\def\Bbb{\twelveDouble}%
\def\LieFont{\twelveSerifItalics}%
\nt\normalbaselines\baselineskip=13pt%
}

\def\AuthorStyle{\parindent=0pt\parskip=0pt\normalbaselineskip=14pt%
\def\nt{\tenSerif}%
\def\rm{\fam0\tenSerif}%
\textfont0=\tenSerif\scriptfont0=\sevenSerif\scriptscriptfont0=\fiveSerif
\textfont1=\tenMath\scriptfont1=\sevenMath\scriptscriptfont1=\fiveMath
\textfont2=\tenSymbols\scriptfont2=\sevenSymbols\scriptscriptfont2=\fiveSymbols
\textfont3=\tenMoreSymbols\scriptfont3=\sevenMoreSymbols\scriptscriptfont3=\fiveMoreSymbols
\textfont\itfam=\tenSerifItalics\def\it{\fam\itfam\tenSerifItalics}%
\textfont\slfam=\tenSerifSlanted\def\sl{\fam\slfam\tenSerifSlanted}%
\textfont\ttfam=\tenTypewriter\def\tt{\fam\ttfam\tenTypewriter}%
\textfont\bffam=\tenSerifBold%
\def\bf{\fam\bffam\tenSerifBold}\scriptfont\bffam=\sevenSerifBold\scriptscriptfont\bffam=\fiveSerifBold%
\def\cal{\tenSymbols}%
\def\greekbold{\tenMathBold}%
\def\gothic{\tenGothic}%
\def\Bbb{\tenDouble}%
\def\LieFont{\tenSerifItalics}%
\nt\normalbaselines\baselineskip=14pt%
}

\def\AddressStyle{\parindent=0pt\parskip=0pt\normalbaselineskip=14pt%
\def\nt{\eightSerif}%
\def\rm{\fam0\eightSerif}%
\textfont0=\eightSerif\scriptfont0=\sevenSerif\scriptscriptfont0=\fiveSerif
\textfont1=\eightMath\scriptfont1=\sevenMath\scriptscriptfont1=\fiveMath
\textfont2=\eightSymbols\scriptfont2=\sevenSymbols\scriptscriptfont2=\fiveSymbols
\textfont3=\eightMoreSymbols\scriptfont3=\sevenMoreSymbols\scriptscriptfont3=\fiveMoreSymbols
\textfont\itfam=\eightSerifItalics\def\it{\fam\itfam\eightSerifItalics}%
\textfont\slfam=\eightSerifSlanted\def\sl{\fam\slfam\eightSerifSlanted}%
\textfont\ttfam=\eightTypewriter\def\tt{\fam\ttfam\eightTypewriter}%
\textfont\bffam=\eightSerifBold%
\def\bf{\fam\bffam\eightSerifBold}\scriptfont\bffam=\sevenSerifBold\scriptscriptfont\bffam=\fiveSerifBold%
\def\cal{\eightSymbols}%
\def\greekbold{\eightMathBold}%
\def\gothic{\eightGothic}%
\def\Bbb{\eightDouble}%
\def\LieFont{\eightSerifItalics}%
\nt\normalbaselines\baselineskip=14pt%
}

\def\AbstractStyle{\parindent=0pt\parskip=0pt\normalbaselineskip=12pt%
\def\nt{\eightSerif}%
\def\rm{\fam0\eightSerif}%
\textfont0=\eightSerif\scriptfont0=\sevenSerif\scriptscriptfont0=\fiveSerif
\textfont1=\eightMath\scriptfont1=\sevenMath\scriptscriptfont1=\fiveMath
\textfont2=\eightSymbols\scriptfont2=\sevenSymbols\scriptscriptfont2=\fiveSymbols
\textfont3=\eightMoreSymbols\scriptfont3=\sevenMoreSymbols\scriptscriptfont3=\fiveMoreSymbols
\textfont\itfam=\eightSerifItalics\def\it{\fam\itfam\eightSerifItalics}%
\textfont\slfam=\eightSerifSlanted\def\sl{\fam\slfam\eightSerifSlanted}%
\textfont\ttfam=\eightTypewriter\def\tt{\fam\ttfam\eightTypewriter}%
\textfont\bffam=\eightSerifBold%
\def\bf{\fam\bffam\eightSerifBold}\scriptfont\bffam=\sevenSerifBold\scriptscriptfont\bffam=\fiveSerifBold%
\def\cal{\eightSymbols}%
\def\greekbold{\eightMathBold}%
\def\gothic{\eightGothic}%
\def\Bbb{\eightDouble}%
\def\LieFont{\eightSerifItalics}%
\nt\normalbaselines\baselineskip=12pt%
}

\def\RefsStyle{\parindent=0pt\parskip=0pt%
\def\nt{\eightSerif}%
\def\rm{\fam0\eightSerif}%
\textfont0=\eightSerif\scriptfont0=\sevenSerif\scriptscriptfont0=\fiveSerif
\textfont1=\eightMath\scriptfont1=\sevenMath\scriptscriptfont1=\fiveMath
\textfont2=\eightSymbols\scriptfont2=\sevenSymbols\scriptscriptfont2=\fiveSymbols
\textfont3=\eightMoreSymbols\scriptfont3=\sevenMoreSymbols\scriptscriptfont3=\fiveMoreSymbols
\textfont\itfam=\eightSerifItalics\def\it{\fam\itfam\eightSerifItalics}%
\textfont\slfam=\eightSerifSlanted\def\sl{\fam\slfam\eightSerifSlanted}%
\textfont\ttfam=\eightTypewriter\def\tt{\fam\ttfam\eightTypewriter}%
\textfont\bffam=\eightSerifBold%
\def\bf{\fam\bffam\eightSerifBold}\scriptfont\bffam=\sevenSerifBold\scriptscriptfont\bffam=\fiveSerifBold%
\def\cal{\eightSymbols}%
\def\greekbold{\eightMathBold}%
\def\gothic{\eightGothic}%
\def\Bbb{\eightDouble}%
\def\LieFont{\eightSerifItalics}%
\nt\normalbaselines\baselineskip=10pt%
}



%
%


\def\ModeYes{yes}
\def\ModeNo{no}

\def\ModeUndef{undefined}


\def\nx{\noexpand}
\def\ni{\noindent}
\def\newpage{\vfill\eject}

\def\ss{\vskip 5pt}
\def\ms{\vskip 10pt}
\def\bs{\vskip 20pt}

 \def\,{\mskip\thinmuskip}
 \def\!{\mskip-\thinmuskip}
 \def\>{\mskip\medmuskip}
 \def\;{\mskip\thickmuskip}

%
% ... 
%

\def\refsModePost{post}
\def\refsModeAuto{auto}

\def\dbRefsSatusModeOk{ok}
\def\dbRefsSatusModeError{error}
\def\dbRefsSatusModeWarning{warning}


\newcount\BNUM
\BNUM=0

\def\refs{}

\def\SetModePost{\xdef\refsMode{\refsModePost}}			
\SetModePost

\def\dbRefsStatusOk{%
	\xdef\dbRefsStatus{\dbRefsSatusModeOk}%
	\xdef\dbRefsError{\ModeNo}%
	\xdef\dbRefsWarning{\ModeNo}%
	\xdef\dbRefsInfo{\ModeNo}%
}

\def\dbRefs{%
}

\def\dbRefsGet#1{%
	\xdef\found{N}\xdef\ikey{#1}\dbRefsStatusOk%
	\xdef\key{\ModeUndef}\xdef\tag{\ModeUndef}\xdef\tail{\ModeUndef}%
	\dbRefs%
}

\def\NextRefsTag{%
	\global\advance\BNUM by 1%
}
\def\ShowTag#1{{\bf [#1]}}

\def\dbRefsInsert#1#2{%
\dbRefsGet{#1}%
\if\found Y %
   \xdef\dbRefsStatus{\dbRefsSatusModeWarning}%
   \xdef\dbRefsWarning{record is already there}%
   \xdef\dbRefsInfo{record not inserted}%
\else%
   \toks2=\expandafter{\dbRefs}%
   \ifx\refsMode\refsModeAuto \NextRefsTag
    \xdef\dbRefs{%
   	\the\toks2 \nx\xdef\nx\dbx{#1}%
	\nx\ifx\nx\ikey %
		\nx\dbx\nx\xdef\nx\found{Y}%
		\nx\xdef\nx\key{#1}%
		\nx\xdef\nx\tag{\the\BNUM}%
		\nx\xdef\nx\tail{#2}%
	\nx\fi}%
	\global\xdef\refs{\refs \ss\ni[\the\BNUM]\ #2\par}
   \fi%
   \ifx\refsMode\refsModePost 
    \xdef\dbRefs{%
   	\the\toks2 \nx\xdef\nx\dbx{#1}%
	\nx\ifx\nx\ikey %
		\nx\dbx\nx\xdef\nx\found{Y}%
		\nx\xdef\nx\key{#1}%
		\nx\xdef\nx\tag{\ModeUndef}%
		\nx\xdef\nx\tail{#2}%
	\nx\fi}%
   \fi%
\fi%
}

\def\dbRefsEdit#1#2#3{\dbRefsGet{#1}%
\if\found N 
   \xdef\dbRefsStatus{\dbRefsSatusModeError}%
   \xdef\dbRefsError{record is not there}%
   \xdef\dbRefsInfo{record not edited}%
\else%
   \toks2=\expandafter{\dbRefs}%
   \xdef\dbRefs{\the\toks2%
   \nx\xdef\nx\dbx{#1}%
   \nx\ifx\nx\ikey\nx\dbx %
	\nx\xdef\nx\found{Y}%
	\nx\xdef\nx\key{#1}%
	\nx\xdef\nx\tag{#2}%
	\nx\xdef\nx\tail{#3}%
   \nx\fi}%
\fi%
}

\def\bib#1#2{\RefsStyle\dbRefsInsert{#1}{#2}%
	\ifx\dbRefsStatus\dbRefsSatusModeWarning %
		\message{^^J}%
		\message{WARNING: Reference [#1] is doubled.^^J}%
	\fi%
}

\def\ref#1{\dbRefsGet{#1}%
\ifx\found N %
  \message{^^J}%
  \message{ERROR: Reference [#1] unknown.^^J}%
  \ShowTag{??}%
\else%
	\ifx\tag\ModeUndef \NextRefsTag%
		\dbRefsEdit{#1}{\the\BNUM}{\tail}%
		\dbRefsGet{#1}%
		\global\xdef\refs{\refs \ss\ni [\tag]\ \tail\par}
	\fi
	\ShowTag{\tag}%
\fi%
}

\def\ShowBiblio{\ms\Ensure{\SectionEnsure}%
{\SectionStyle\ni References}%
{\RefsStyle\refs}%
}

\newcount\CHANGES
\CHANGES=0
\def\AuxFile{7}
\def\PreventDoubleOn{\xdef\PreventDoubleLabel{\ModeYes}}

\PreventDoubleOn

\def\StoreLabel#1#2{\xdef\itag{#2}
 \ifx\PreModeStatus\ModeNo %
   \message{^^J}%
   \errmessage{You can't use Check without starting with OpenPreMode (and finishing with ClosePreMode)^^J}%
 \else%
   \immediate\write\AuxFile{\nx\dbLabelPreInsert{#1}{\itag}}%
   \dbLabelGet{#1}%
   \ifx\itag\tag %
   \else%
	\global\advance\CHANGES by 1%
 	\xdef\itag{(?.??)}%
    \fi%
   \fi%
}

\def\PreModeStatus{\ModeNo}

\def\ClosePreMode{\immediate\closeout\AuxFile%
  \ifnum\CHANGES=0%
	\message{^^J}%
	\message{**********************************^^J}%
	\message{**  NO CHANGES TO THE AuxFile  **^^J}%
	\message{**********************************^^J}%
 \else%
	\message{^^J}%
	\message{**************************************************^^J}%
	\message{**  PLAEASE TYPESET IT AGAIN (\the\CHANGES)  **^^J}%
    \errmessage{**************************************************^^ J}%
  \fi%
  \edef\PreModeStatus{\ModeNo}%
}

\def\dbLabelSatusModeOk{ok}

\def\dbLabelSatusModeWarning{warning}

\def\dbLabelStatusOk{%
	\xdef\dbLabelStatus{\dbLabelSatusModeOk}%
	\xdef\dbLabelError{\ModeNo}%
	\xdef\dbLabelWarning{\ModeNo}%
	\xdef\dbLabelInfo{\ModeNo}%
}

\def\dbLabel{%
}

\def\dbLabelGet#1{%
	\xdef\found{N}\xdef\ikey{#1}\dbLabelStatusOk%
	\xdef\key{\ModeUndef}\xdef\tag{\ModeUndef}\xdef\pre{\ModeUndef}%
	\dbLabel%
}

\def\ShowLabel#1{%
 \dbLabelGet{#1}%
 \ifx\tag \ModeUndef %
 	\global\advance\CHANGES by 1%
 	(?.??)%
 \else%
 	\tag%
 \fi%
}

\def\dbLabelPreInsert#1#2{\dbLabelGet{#1}%
\if\found Y %
  \xdef\dbLabelStatus{\dbLabelSatusModeWarning}%
   \xdef\dbLabelWarning{Label is already there}%
   \xdef\dbLabelInfo{Label not inserted}%
   \message{^^J}%
   \errmessage{Double pre definition of label [#1]^^J}%
\else%
   \toks2=\expandafter{\dbLabel}%
    \xdef\dbLabel{%
   	\the\toks2 \nx\xdef\nx\dbx{#1}%
	\nx\ifx\nx\ikey %
		\nx\dbx\nx\xdef\nx\found{Y}%
		\nx\xdef\nx\key{#1}%
		\nx\xdef\nx\tag{#2}%
		\nx\xdef\nx\pre{\ModeYes}%
	\nx\fi}%
\fi%
}

\def\dbLabelInsert#1#2{\dbLabelGet{#1}%
\xdef\itag{#2}%
\dbLabelGet{#1}%
\if\found Y %
	\ifx\tag\itag %
	\else%
	   \ifx\PreventDoubleLabel\ModeYes %
		\message{^^J}%
		\errmessage{Double definition of label [#1]^^J}%
	   \else%
		\message{^^J}%
		\message{Double definition of label [#1]^^J}%
	   \fi%
	\fi%
   \xdef\dbLabelStatus{\dbLabelSatusModeWarning}%
   \xdef\dbLabelWarning{Label is already there}%
   \xdef\dbLabelInfo{Label not inserted}%
\else%
   \toks2=\expandafter{\dbLabel}%
    \xdef\dbLabel{%
   	\the\toks2 \nx\xdef\nx\dbx{#1}%
	\nx\ifx\nx\ikey %
		\nx\dbx\nx\xdef\nx\found{Y}%
		\nx\xdef\nx\key{#1}%
		\nx\xdef\nx\tag{#2}%
		\nx\xdef\nx\pre{\ModeNo}%
	\nx\fi}%
\fi%
}


\newcount\PART
\newcount\CHAPTER
\newcount\SECTION
\newcount\SUBSECTION
\newcount\FNUMBER

\PART=0
\CHAPTER=0
\SECTION=0
\SUBSECTION=0	
\FNUMBER=0

\def\LastPart{\ModeUndef}
\def\LastChapter{\ModeUndef}
\def\LastSection{\ModeUndef}
\def\LastSubSection{\ModeUndef}
\def\LastClaim{\ModeUndef}
\def\Last{\ModeUndef}

\newdimen\TOBOTTOM
\newdimen\LIMIT

\def\Ensure#1{\ \par\ \immediate\LIMIT=#1\immediate\TOBOTTOM=\the\pagegoal\advance\TOBOTTOM by -\pagetotal%
\ifdim\TOBOTTOM<\LIMIT\newpage \else%
\vskip-\parskip\vskip-\parskip\vskip-\baselineskip\fi}

\def\PartLabel{\the\PART}
\def\NewPart#1{\global\advance\PART by 1%
         \bs\ni{\PartStyle  Part \PartLabel:}
         \bs\ni{\PartStyle #1}\newpage%
         \CHAPTER=0\SECTION=0\SUBSECTION=0\FNUMBER=0%
         \gdef\Left{#1}%
         \global\edef\Last{\PartLabel}%
         \global\edef\LastPart{\PartLabel}%
         \global\edef\LastChapter{\ModeUndef}%
         \global\edef\LastSection{\ModeUndef}%
         \global\edef\LastSubSection{\ModeUndef}%
         \global\edef\LastClaim{\ModeUndef}}
\def\ChapterLabel{\the\CHAPTER}
\def\NewChapter#1{\global\advance\CHAPTER by 1%
         \bs\ni{\ChapterStyle  Chapter \ChapterLabel: #1}\ms%
         \SECTION=0\SUBSECTION=0\FNUMBER=0%
         \gdef\Left{#1}%
         \global\edef\Last{\ChapterLabel}%
         \global\edef\LastChapter{\ChapterLabel}%
         \global\edef\LastSection{\ModeUndef}%
         \global\edef\LastSubSection{\ModeUndef}%
         \global\edef\LastClaim{\ModeUndef}}
\def\SectionEnsure{3cm}
\def\NewSection#1{\Ensure{\SectionEnsure}\gdef\SectionLabel{\the\SECTION}\global\advance\SECTION by 1%
         \ms\ni{\SectionStyle  \SectionLabel.\ #1}\ss%
         \SUBSECTION=0\FNUMBER=0%
         \gdef\Left{#1}%
         \global\edef\Last{\SectionLabel}%
         \global\edef\LastSection{\SectionLabel}%
         \global\edef\LastSubSection{\ModeUndef}%
         \global\edef\LastClaim{\ModeUndef}}
\def\NewAppendix#1#2{\Ensure{\SectionEnsure}\gdef\SectionLabel{#1}\global\advance\SECTION by 1%
         \bs\ni{\SectionStyle  Appendix \SectionLabel.\ #2}\ss%
         \SUBSECTION=0\FNUMBER=0%
         \gdef\Left{#2}%
         \global\edef\Last{\SectionLabel}%
         \global\edef\LastSection{\SectionLabel}%
         \global\edef\LastSubSection{\ModeUndef}%
         \global\edef\LastClaim{\ModeUndef}}
\def\Acknowledgements{\Ensure{\SectionEnsure}\gdef\SectionLabel{}%
         \ms\ni{\SectionStyle  Acknowledgments}\ss%
         \SECTION=0\SUBSECTION=0\FNUMBER=0%
         \gdef\Left{}%
         \global\edef\Last{\ModeUndef}%
         \global\edef\LastSection{\ModeUndef}%
         \global\edef\LastSubSection{\ModeUndef}%
         \global\edef\LastClaim{\ModeUndef}}
\def\SubSectionEnsure{2cm}
\def\SubSectionLabel{\ifnum\SECTION>0 \the\SECTION.\fi\the\SUBSECTION}
\def\NewSubSection#1{\Ensure{\SubSectionEnsure}\global\advance\SUBSECTION by 1%
         \ms\ni{\SubSectionStyle #1}\ss%
         \global\edef\Last{\SubSectionLabel}%
         \global\edef\LastSubSection{\SubSectionLabel}}
\def\SetNumberingModeN{\def\ClaimLabel{(\the\FNUMBER)}}
\def\SetNumberingModeSN{\def\ClaimLabel{(\ifnum\SECTION>0 \SectionLabel.\fi%
      \the\FNUMBER)}}
\def\SetNumberingModeCSN{\def\ClaimLabel{(\ifnum\CHAPTER>0 \the\CHAPTER.\fi%
      \ifnum\SECTION>0 \SectionLabel.\fi%
      \the\FNUMBER)}}

\def\NewClaim{\global\advance\FNUMBER by 1%
    \ClaimLabel%
    \global\edef\LastClaim{\ClaimLabel}%
    \global\edef\Last{\ClaimLabel}}

\def\HideLabels{\xdef\ShowLabelsMode{\ModeNo}}
\HideLabels

\def\fn{\eqno{\NewClaim}} 
\def\fl#1{%
\ifx\ShowLabelsMode\ModeYes%
 \eqno{{\buildrel{\hbox{\AbstractStyle[#1]}}\over{\hfill\NewClaim}}}%
\else%
 \eqno{\NewClaim}%
\fi%
\dbLabelInsert{#1}{\ClaimLabel}}
\def\fprel#1{\global\advance\FNUMBER by 1\StoreLabel{#1}{\ClaimLabel}%
\ifx\ShowLabelsMode\ModeYes%
\eqno{{\buildrel{\hbox{\AbstractStyle[#1]}}\over{\hfill.\itag}}}%
\else%
 \eqno{\itag}%
\fi%
}

\def\cl#1{\global\advance\FNUMBER by 1\dbLabelInsert{#1}{\ClaimLabel}%
\ifx\ShowLabelsMode\ModeYes%
${\buildrel{\hbox{\AbstractStyle[#1]}}\over{\hfill\ClaimLabel}}$%
\else%
  $\ClaimLabel$%
\fi%
}
\def\cprel#1{\global\advance\FNUMBER by 1\StoreLabel{#1}{\ClaimLabel}%
\ifx\ShowLabelsMode\ModeYes%
${\buildrel{\hbox{\AbstractStyle[#1]}}\over{\hfill.\itag}}$%
\else%
  $\itag$%
\fi%
}

\def\Note{\ms\leftskip 3cm\rightskip 1.5cm\AbstractStyle}
\def\endNote{\par\leftskip 2cm\rightskip 0cm\NormalStyle\ss}


\parindent=7pt
\leftskip=2cm
\newcount\SideIndent
\newcount\SideIndentTemp
\SideIndent=0
\newdimen\SectionIndent
\SectionIndent=-8pt

\def\sidebar{\vrule height15pt width.2pt }
\def\endcorner{\hbox{\hbox{\vrule height6pt width.2pt}\vbox to6pt{\vfill\hbox
to4pt{\leaders\hrule height0.2pt\hfill}}}}
\def\begincorner{\hbox{\hbox{\vrule height6pt width.2pt}\vbox to6pt{\hbox
to4pt{\leaders\hrule height0.2pt\hfill}}}}
\def\endbegincorner{\hbox{\vbox to15pt{\endcorner\vskip-6pt\begincorner\vfill}}}
\def\SideShow{\SideIndentTemp=\SideIndent \ifnum \SideIndentTemp>0 
\loop\sidebar\hskip 2pt \advance\SideIndentTemp by-1\ifnum \SideIndentTemp>1 \repeat\fi}

\def\BeginSection{{\vbadness 100000 \par\ni\hskip\SectionIndent%
\SideShow\vbox to 15pt{\vfill\begincorner}}\global\advance\SideIndent by1\vskip-10pt}

\def\EndSection{{\vbadness 100000 \par\ni\global\advance\SideIndent by-1%
\hskip\SectionIndent\SideShow\vbox to15pt{\endcorner\vfill}\vskip-10pt}}

\def\EndBeginSection{{\vbadness 100000\par\ni%
\global\advance\SideIndent by-1\hskip\SectionIndent\SideShow
\vbox to15pt{\vfill\endbegincorner}}%
\global\advance\SideIndent by1\vskip-10pt}

\def\ShowBeginCorners#1{%
\SideIndentTemp =#1 \advance\SideIndentTemp by-1%
\ifnum \SideIndentTemp>0 %
\vskip-15truept\hbox{\kern 2truept\vbox{\hbox{\begincorner}%
\ShowBeginCorners{\SideIndentTemp}\vskip-3truept}}%
\fi%
}

\def\ShowEndCorners#1{%
\SideIndentTemp =#1 \advance\SideIndentTemp by-1%
\ifnum \SideIndentTemp>0 %
\vskip-15truept\hbox{\kern 2truept\vbox{\hbox{\endcorner}%
\ShowEndCorners{\SideIndentTemp}\vskip 2truept}}%
\fi%
}

\def\BeginSections#1{{\vbadness 100000 \par\ni\hskip\SectionIndent%
\SideShow\vbox to 15pt{\vfill\ShowBeginCorners{#1}}}\global\advance\SideIndent by#1\vskip-10pt}

\def\EndSections#1{{\vbadness 100000 \par\ni\global\advance\SideIndent by-#1%
\hskip\SectionIndent\SideShow\vbox to15pt{\vskip15pt\ShowEndCorners{#1}\vfill}\vskip-10pt}}

\def\EndBeginSections#1#2{{\vbadness 100000\par\ni%
\global\advance\SideIndent by-#1%
\hbox{\hskip\SectionIndent\SideShow\kern-2pt%
\vbox to15pt{\vskip15pt\ShowEndCorners{#1}\vskip4pt\ShowBeginCorners{#2}}}}%
\global\advance\SideIndent by#2\vskip-10pt}




%
%


\def\al{\alpha}
\def\be{\beta}
\def\de{\delta}
\def\ga{\gamma}

\def\ep{\epsilon}

\def\la{\lambda}

\def\om{\omega}
\def\si{\sigma}

\def\ka{\kappa}

\def\Ga{\Gamma}

\def\La{\Lambda}

 
 \def\calU{{\hbox{\cal U}}}
 
 \def\calC{{\hbox{\cal C}}}

 \def\calL{{\hbox{\cal L}}}
 \def\calE{{\hbox{\cal E}}}



 \def\one{{\hbox{\Bbb I}}}


\def\Div{{\hbox{Div}}}

\def\Spin{{\hbox{Spin}}}
\def\SO{{\hbox{SO}}}

\def\GL{{\hbox{GL}}}
\def\det{{\hbox{det}}}

\def\ip{\hbox to4pt{\leaders\hrule height0.3pt\hfill}\vbox to8pt{\leaders\vrule width0.3pt\vfill}\kern 2pt}
\def\QDE{\hfill\hbox{\ }\vrule height4pt width4pt depth0pt} 
\def\del{\partial}
\def\na{\nabla}

\def\Lie{\hbox{\LieFont \$}}

\def\arr{\rightarrow}

\def\then{\Rightarrow}

%
%

\def\cases#1{\left\{\eqalign{#1}\right.}
\NormalStyle
\SetNumberingModeSN
\PreventDoubleOn

\long\def\title#1{\centerline{\TitleStyle\ni#1}}

\long\def\author#1{\ms\centerline{\AuthorStyle by {\it #1}}}

\long\def\address#1{\ss\centerline{\AddressStyle #1}\par}
\long\def\moreaddress#1{\centerline{\AddressStyle #1}\par}
\def\abstract{\ms\leftskip 3cm\rightskip .5cm\AbstractStyle{\bf \ni Abstract:}\ }
\def\endabstract{\par\leftskip 2cm\rightskip 0cm\NormalStyle\ss}

\SetNumberingModeSN

\def\frac[#1/#2]{\hbox{$#1\over#2$}}
\def\Frac[#1/#2]{{#1\over#2}}
\def\({\left(}
\def\){\right)}
\def\[{\left[}
\def\]{\right]}
\def\^#1{{}^{#1}_{\>\cdot}}
\def\_#1{{}_{#1}^{\>\cdot}}
\def\Label=#1{{\buildrel {\hbox{\fiveSerif \ShowLabel{#1}}}\over =}}
\def\<{\kern -1pt}

\def\sign{\hbox{sign}}


\def\ExpandAllCNotes{\long\def\CNote##1{%
\BeginSection
	\Note%
 		##1%
	\endNote%
\EndSection%
}}
\ExpandAllCNotes
%
%
%
%


\def\frame#1{\vbox{\hrule\hbox{\vrule\vbox{\kern2pt\hbox{\kern2pt#1\kern2pt}\kern2pt}\vrule}\hrule\kern-4pt}}

\def\Box to #1#2#3{\frame{\vtop{\hbox to #1{\hfill #2 \hfill}\hbox to #1{\hfill #3 \hfill}}}}



\def\ubal{\underline{\al}\kern1pt}
\def\obal{\overline{\al}\kern1pt}

\def\ubR{\underline{R}\kern1pt}
\def\obR{\overline{R}\kern1pt}
\def\ubom{\underline{\om}\kern1pt}
\def\obxi{\overline{\xi}\kern1pt}
\def\ubu{\underline{u}\kern1pt}
\def\ube{\underline{e}\kern1pt}
\def\obe{\overline{e}\kern1pt}

\def\beR{{}^\be\<\<R}

\bib{Universality}{A. Borowiec, M. Ferraris, M. Francaviglia, I. Volovich,
{\it Universality of Einstein Equations for the Ricci Squared Lagrangians},
Class. Quantum Grav. 15, 43-55, 1998}

\bib{Faraoni}{V.Faraoni, 
{\it f(R) gravity: successes and challenges},
arXiv:0810.2602}

\bib{NoGo}{E.Barausse, T.P. Sotiriou, J.C.Miller,
{\it A no-go theorem for polytropic spheres in Palatini $f(R)$ gravity},  
Classical Quantum Gravity {\bf 25}, 2008 
}

\bib{Olmo}{G.J. Olmo, 
{\it Re-examination of polytropic spheres in Palatini f(R) gravity}, 
Phys.Rev.D78, 2008}

\bib{Capozziello}{S. Capozziello, V.F. Cardone, V. Salzano,
{\it Cosmography of $f(R)$ gravity},
Phys.Rev.D{\bf 78}, 063504, 2008
}

\bib{DM}{S. Capozziello, M. De Laurentis, M. Francaviglia, S. Mercadante,
{\it From Dark Energy and Dark Matter to Dark Metric},
arXiv:0805.3642
}

\bib{Allemandi}{G. Allemandi, A. Borowiec, M. Francaviglia, S.D. Odintsov,
{\it Dark Energy Dominance and Cosmic Acceleration in First Order Formalism},
Phys.Rev.D{\bf 72}, 063505, 2005
}

\bib{Odintzov}{S. Nojiri, S.D. Odintsov,
{\it Modified gravity as realistic candidate for dark energy, inflation and dark matter}
AIP Conf.Proc. 1115, 2009, 212-217;
arXiv:0810.1557}

\bib{Magnano}{G. Magnano, M. Ferraris, M. Francaviglia, 
{\it Nonlinear gravitational Lagrangians},
Gen.Rel.Grav. {\bf 19}(5), 1987, 465-479}

\bib{Faraoni2}{T.P. Sotiriou, V.Faraoni, 
{\it $f(R)$ theories of gravity},
arXiv:0805.1726}

\bib{Wald}{R.M. Wald,
{\it General Relativity}
The University of Chicago Press, 1984}

\bib{Barbero}{F.\ Barbero, 
{\it Real Ashtekar variables for Lorentzian signature space-time},
Phys.\ Rev.\ {\it D51}, 5507, 1996}

\bib{Immirzi}{G.\ Immirzi, 
{\it Quantum Gravity and Regge Calculus},
Nucl.\ Phys.\ Proc.\ Suppl.\ {\bf 57}, 65-72}

\bib{AshtekarLewandowski}{A.\ Ashtekar, J.\ Lewandowski,
{\it Background Independent Quantum Gravity: a Status Report}, 
gr-qc/0404018}

\bib{Holst}{S.\ Holst, 
{\it Barbero's Hamiltonian Derived from a Generalized Hilbert-Palatini Action},
Phys.\ Rev.\ {\bf D53}, 5966, 1996}

\bib{myBarbero}{L. Fatibene, M.Francaviglia, C.Rovelli, 
{\it On a Covariant Formulation of the Barberi-Immirzi Connection}
CQG 24 (2007) 3055-3066; gr-qc/0702134v1}

\bib{myHolst}{L. Fatibene, M.Francaviglia, C.Rovelli, 
{\it Spacetime Lagrangian Formulation of Barbero-Immirzi Gravity} 
CQG 24 (2007) 4207-4217; gr-qc/0706.1899}

\bib{Libro}{L.\ Fatibene, M.\ Francaviglia, 
{\it Natural and gauge natural formalism for classical field theories. A geometric perspective including spinors and gauge theories}, 
Kluwer Academic Publishers, Dordrecht, 2003}

\bib{Rovelli}{C.\ Rovelli,
{\it Quantum Gravity}, 
Cambridge University Press, Cambridge, 2004}

\bib{Bojowald}{M. Bojowald,
{\it Consistent Loop Quantum Cosmology}
Class.Quant.Grav.{\bf 26} 075020, 2009}

\bib{Thieman}{T. Thiemann,
{\it LoopQuantumGravity: An InsideView}, 
hep-th/0608210}

\bib{Augmented}{L.~Fatibene, M.~Ferraris, M.~Francaviglia,
{\it Augmented Variational Principles and Relative Conservation Laws in Classical Field Theory},
Int. J. Geom. Methods Mod. Phys., {\bf 2}(3), (2005), pp. 373-392; [math-ph/0411029v1]}

\bib{HolstCQ}{L.Fatibene, M.Ferraris, M.Francaviglia, G.Pacchiella, 
{\it Entropy of SelfÐGravitating Systems from HolstÕs Lagrangian}, 
Int. Journal of Geometrical Methods in Modern Physics,  {\bf 6}(2), 2009; gr-qc/0808.3845v2
}

 \bib{KL}{L. Fatibene, M. Ferraris, M. Francaviglia, M. Godina,
in: Proceedings of {\it ``6th International Conference on Differential Geometry
and its Applications, August 28--September 1, 1995"}, (Brno, Czech Republic),
Editor: I. Kol{\'a}{\v r}, MU University, Brno, Czech Republic (1996) 549.}

\bib{CovariantFirstOrderTheory}{M. Ferraris, M. Francaviglia,
{\it Covariant first-order Lagrangians, energy-density and superpotentials in general relativity},
Gen.Rel.Grav. {\bf 22}(9), 1990, 965-985}

\bib{myHOT}{L. Fatibene, M.Francaviglia, S.Mercadante, 
{\it About Boundary Terms in Higher Order Theories},
(in preparation)
}

\bib{MM}{S. Capozziello, M.F. De Laurentis, M. Francaviglia, S. Mercadante,
{\it From Dark Energy \& Dark Matter to Dark Metric},
Foundations of Physics 39 (2009) 1161-1176
gr-qc/0805.3642v4}

\bib{CoV}{I.M. Gelfand, S.V. Fomin, {\it Calculus of Variations},
Prentice-Hall Inc., (1963)}

\NormalStyle

\title{New Cases of Universality Theorem for Gravitational Theories\footnote{$^*$}{{\AbstractStyle
	This paper is published despite the effects of the Italian law 133/08 ({\tt http://groups.google.it/group/scienceaction}). 
        This law drastically reduces public funds to public Italian universities, which is particularly dangerous for free scientific research, 
        and it will prevent young researchers from getting a position, either temporary or tenured, in Italy.
        The authors are protesting against this law to obtain its cancellation.\goodbreak}}}

\author{L.Fatibene$^{a, b}$, M.Ferraris$^a$, M.Francaviglia$^{a,b,c}$}

\address{$^a$ Department of Mathematics, University of Torino (Italy)}

\moreaddress{$^b$ INFN - Iniziativa Specifica Na12}

\moreaddress{$^c$ LCS, University of Calabria (Italy)}

\abstract
The ``Universality Theorem'' for gravity shows that  $f(R)$ theories (in their metric-affine formulation) in vacuum  are dynamically equivalent 
to vacuum Einstein equations with suitable cosmological constants. This holds true for a generic (i.e.~except sporadic degenerate cases) 
analytic function $f(R)$ and standard gravity without cosmological constant is reproduced if $f$ is the identity function (i.e.~$f(R)=R$).

The theorem is here extended introducing  in dimension $4$ a $1$-parameter family of invariants $\beR$ inspired by the Barbero-Immirzi formulation of GR
(which in the Euclidean sector includes also selfdual formulation).
It will be proven that $f(\beR)$ theories so defined are dynamically equivalent to the corresponding metric--affine $f(R)$ theory.
In particular for the function $f(R)=R$ the standard equivalence between GR and Holst Lagrangian is obtained.
\endabstract

\NewSection{Introduction}

It is well-known that for almost any analytic function $f(R)$ the metric-affine theory with Lagrangian $L_f (g, j^1\Ga)=\sqrt{g} f(R(g, j^1\Ga))$ is dynamically equivalent to standard GR with a suitably quantized cosmological constant (encoded by the function $f$); see \ref{Universality} and \ref{Magnano}.
Hereafter $j^1\Ga$ refers to the fact that the Lagrangian depends on the connection $\Ga$ and its first derivatives.

The original universality theorem was established in vacuum in \ref{Universality}. In fact, matter coupling produces in $f(R)$ theories new effects with respect to Einstein equations, since a new metric conformal to the original one can be defined (see \ref{MM}).
The theory in these new variables is quite similar to standard GR though with the addition of an effective energy momentum tensor; see \ref{Magnano}.
Such new effects have been recently investigated aiming to find a specific $f(R)$ theory able to model dark energy and dark matter phenomenology; 
see e.g.~\ref{Allemandi}, \ref{Capozziello}, \ref{Odintzov}, \ref{DM}. 

The metric-affine formulation has been recently  criticized. The most serious criticism is based on a theorem which shows that,
under specific state equation hypotheses, internal and external solutions
for a (stationary, spherically symmetric) polytropic star do not match on the boundary and produce singularities on the star surface; see \ref{NoGo}.
However, as it often happens when precise no-go theorems are formulated, it has been later argued that the hypotheses of this theorem 
are in fact physically unreasonable; see \ref{Olmo}.
In this particular case it has been shown that (at least for specific examples of $f$) the singularity depends on matter densities far lower 
than the one that is reasonably expected. We shall not discuss further these aspects here. 

Another criticism to $f(R)$ theories (this time in purely-metric formulation; see \ref{Magnano}) is based on fixing the variation of first derivatives of field on the boundary; see \ref{Faraoni2}.
This criticism seems to be based on a physically and also mathematically questionable method discussed  \ref{Wald}, 
that is meaningful only in standard GR;
in standard purely-metric GR, one can in fact subtract a suitable boundary term from the action so that fixing first derivatives of the metric 
on the boundary (as it is prescribed by Calculus of Variations; \ref{CoV}) is not necessary any longer provided one accepts to modify the Hilbert Lagrangian by
adding suitable {\it ad-hoc} boundary counterterms.
 The same unconventional procedure, however,  cannot be carried over for a generic $f(R)$.
We argue that one cannot consider this a problem; see \ref{myHOT} for a detailed criticism of the method. 
In standard (higher order) variational calculus all derivatives of variations of fields are fixed up to one order less with respect to the effective order of the theory. 
The fact that first derivatives do not need to be fixed in GR, which is a second order gravitational theory when formulated in terms of the Hilbert Lagrangian, is 
just  a consequence of the well-known fact that GR is in fact degenerate and can be formulated even covariantly as an equivalent  first order theory; see \ref{CovariantFirstOrderTheory}.
Moreover, not fixing higher order variations causes unreasonable results in many cases, but we
shall not discuss these aspects here, either;  see \ref{myHOT}. 

We shall here establish the dynamical equivalence between $f(R)$ theory {\it \`a la} Palatini and its Barbero-Immirzi formulation.
The special case for $f(R)=R$ is already well-known; see \ref{Barbero}, \ref{Immirzi}, \ref{Holst}, \ref{myBarbero}, \ref{myHolst}.
The extension to a generic $f(R)$ is new, to the best of our knowledge.

\NewSection{Notation and Holst Formulation of GR }

Holst formulation of GR is the classical basis of LQG formulation in terms of the Barbero-Immirzi connection.

Let us consider a $4$ dimensional (paracompact, connected, orientable) manifold $M$ which allows metrics in signature $\eta=(3,1)$ and global spin structures (i.e.~with zero $1st$ and $2nd$ Stiefel-Whitney classes).
Let $P$ be a principal bundle over $M$ with $\SO(\eta)$ as structure group. 
Notation follows \ref{Libro}.

Let us denote by $J^1P$ the first jet prolongation of $P$; there is a standard right action of $\SO(\eta)$ on $J^1P$, namely the prolongation of the canonical right action $R_g:P\arr P$ on $P$; see \ref{Libro}.
The bundle $C(P)=J^1P/\SO(\eta)$ admits a global family of local coordinates $(x^\mu, \Ga^{ab}_\mu)$ and (global) sections are by definition (global) $\SO(\eta)$-connections
on $P$.  

Let $\la:	\(\GL(4)\times \SO(\eta)\)\times \GL(4)\arr \GL(4): (J, \ell, e)\mapsto J \cdot e\cdot \ell^{-1}$ be the natural action of the group $\GL(4)\times \SO(\eta)$
on the manifold $\GL(4)$. The associated bundle $e(P)= (L(M) \times P) \times_\la \GL(4)$ has coordinates $(x^\mu, e_a^\mu)$
and (global) sections, that always exist under our hypotheses on $M$, are by definition {\it (global) frames}. 

\CNote{
Here {\it frame} is meant in a non-standard sense. A global frame for us is not a global section of the frame bundle $L(M)$, that may not exist, but rather a global section of $e(P)$.
A global frame is here  a family of local sections of $L(M)$ defined on an open covering of $M$ and such that transition functions are valued in $\SO(\eta)$
rather than in $\{\one\}$ as required for global sections of $L(M)$.

This is done since generic manifolds do not allow global sections of $L(M)$; the topological condition for this is physically too strong in general, 
while $e(P)$ always allows global frames in the above hypotheses; see \ref{Libro}.

Moreover, the standard covariant derivative of sections of $e(P)$ is the covariant derivative introduced {\it ad hoc} and used in literature for frames, which is not the standard covariant derivative for sections of $L(M)$; see \ref{Libro}.
}

{\it Vielbein} $e^a_\mu$ are frame inverses; vielbein induce a metric on $M$
$$
g_{\mu\nu}= e^a_\mu \>\eta_{ab}\> e^b_\nu
\fn$$
where $\eta^{ab}$ is the standard diagonal matrix of signature $\eta=(3,1)$.  Hereafter Greek indices are moved up and down by the metric $g_{\mu\nu}$
while Latin indices are moved by $\eta_{ab}$.

Let us now define the curvature $2$-form
$$
R^{ab}=\frac[1/2] R^{ab}{}_{\mu\nu} \> dx^\mu\land dx^\nu
\fn$$
where $ R^{ab}{}_{\mu\nu}(j^1\Ga)$ is the curvature tensor of $\Ga^{ab}_\mu$, the local representative of an arbitrary global 
section $\Ga$ of  $C(P)$. The {\it vielbein form} is the $1$-form
$$
e^a=e^a_\mu \> dx^\mu
\fn$$ 

For the Holst formulation (see \ref{Holst}, \ref{myHolst}, \ref{AshtekarLewandowski}) 
let us set $\calC=C(P)\times_M e(P)$ for the configuration bundle; the Lagrangian is:
$$
L_H= \frac[1/4\ka] \( R^{ab}\land e^c\land e^d \ep_{abcd} - \frac[2/\ga] R^{ab}\land e_a\land e_b\)
\fn$$
where $\ka$ and $\ga$ are constants.
For later covanience this Lagrangian can be written as
$$
\eqalign{
L_H=& \frac[1/8\ka] \( R^{ab}{}_{\mu\nu} e^c_\rho e^d_\si \ep_{abcd} - \frac[2/\ga] R^{ab}{}_{\mu\nu} e_{a\rho} e_{b\si}\)\ep^{\mu\nu\rho\si}\> ds=\cr
=&\frac[e/2\ka] \(  R^{ab}{}_{\mu\nu}  e_a^\mu e_b^\nu   - \frac[1/2\ga] R^{ab}{}_{\mu\nu}   e_c^\mu e_d^\nu \ep^{cd}\_a\_b \) \> ds
= \frac[e/2\ka]  \>\beR\> ds\cr
}
\fl{Holst}$$
where $ds$ is the standard local volume form induced by coordinates, $e$ is the (module of) the determinant of the vielbein $e^a_\mu$ and we set $\be=- \frac[1/2\ga]$ and
$$
\beR= R^{ab}{}_{\mu\nu}  e_a^\mu e_b^\nu   +\be R^{ab}{}_{\mu\nu}   e^{c\mu} e^{d\nu} \ep_{cdab}
\fn$$
Let us also set $R:=R^{ab}{}_{\mu\nu}  e_a^\mu e_b^\nu$ and
$R^b_\nu:=R^{ab}{}_{\mu\nu}  e_a^\mu $ which are functions of $(e, j^1\Ga)$.

Let us stress that the additional term $R^{ab}{}_{\mu\nu}  e^{c\mu} e^{d\nu} \ep_{cdab}$ in $\beR$ is peculiar of the Palatini formulation; 
if we assumed the connection to be metric from the beginning, the curvature would be the Riemann tensor of a metric and the additional term would vanish identically
due to the symmetry properties of Riemann tensors, in particular the first Bianchi identity $R^\al{}_{[\be\mu\nu]}=0$.
One obtains the same result also using no frames, but allowing torsion;
it is sufficient to use the correction $\ep^{\mu\nu\rho\si} R_{\mu\nu\rho\si} (j^1\Ga)$ which does not vanish identically if torsion is allowed.
Here we have chosen to use frames in view of possible applications to LQG or possible applications with spinor couplings.

We refer to \ref{Holst} or \ref{AshtekarLewandowski} for the equivalence between the Lagrangian \ShowLabel{Holst} and standard GR; here it will follow from the universality result proved below, in the particular case $f(\beR)=\>\beR$.

Before proceeding to consider these further extended gravitational theories let us first briefly review the standard results about metric-affine $f(R)$ theories.
The metric-affine formulation of $f(R)$ models is described by the following Lagrangian:
$$
L_f (g, j^1\Ga, j^1\phi)= \sqrt{g} f(R) + \calL_{mat} (g, j^1\phi)
\fl{MetricAffineLag}$$
where $\Ga$ is now a (torsionless) linear connection and $\phi$ is a set of matter fields; the matter Lagrangian does not depend on $\Ga$ (though it could depend on the Levi-Civita connection of the metric $g$). 

The variation of the Lagrangian $L_f$  with respect to the metric $g$ and the connection $\Ga$ is:
$$
\eqalign{
\de L_f=& \sqrt{g}\( - \frac[1/2]f g_{\mu\nu} \de g^{\mu\nu}  + f' \(\de g^{\mu\nu} R_{\mu\nu} + g^{\mu\nu} \de R_{\mu\nu}  \) \) -\sqrt{g}T_{\mu\nu} \de g^{\mu\nu}  =\cr
=&\sqrt{g}\( -\frac[1/2]f  g_{\mu\nu} \de g^{\mu\nu}  + f' \(\de g^{\mu\nu} R_{\mu\nu} + 2 g^{\mu\nu} \na_\la \de u^\la_{\mu\nu}  \) \) -\sqrt{g}T_{\mu\nu} \de g^{\mu\nu} =\cr
=&\sqrt{g}\( f'R_{\mu\nu}  -\frac[1/2] f g_{\mu\nu}   -T_{\mu\nu}  \)\de g^{\mu\nu} 
- 2\na_\la \(\sqrt{g}  f'  g^{\mu\nu}\)  \de u^\la_{\mu\nu}  
 + \na_\la\(2\sqrt{g}  f'  g^{\mu\nu}  \de u^\la_{\mu\nu} \) 
} 
\fn$$
where  $f$ and $f'=\frac[df/dR]$ are understood to be evaluated at $R$ and we set 
$u^\la_{\mu\nu}= \Ga^\la_{\mu\nu} -\de^\la_{(\mu} \Ga^\al_{\nu)\al}$
and $-\sqrt{g}T_{\mu\nu} =\frac[\de L_{mat}/\de g^{\mu\nu}]$.
One should also consider variations with respect to matter fields, which account for matter field equations; we are not interested here in that.

As usual the last boundary term in the variation vanishes because of boundary conditions ($\de u^\al_{\be\mu}=0$ since $\de \Ga^\al_{\be\mu}=0$) and field equations are
$$
\cases{
& f'R_{(\mu\nu)}  -\frac[1/2] f g_{\mu\nu}   = T_{\mu\nu}\cr
&\na_\la \(\sqrt{g}  f'  g^{\mu\nu}\) =0 \cr
}
\fl{FE1}$$
By tracing the first equation with $g$,  one obtains the {\it master equation} $f'(R) R -2 f(R)    =  T$, being $T=g^{\mu\nu}T_{\mu\nu}$; 
except in degenerate cases this can be solved for $R(T)$.

Let us now set $\si=\sign(f')$ and define a new metric $\tilde g_{\mu\nu} = |f'| g_{\mu\nu}$; accordingly one has $\si\sqrt{\tilde g} \tilde g^{\mu\nu}= \sqrt{g}  f' g^{\mu\nu}$. This can be used in the second field equation which implies that $\Ga^\la_{\mu\nu}$ are nothing but the Christoffel symbols of the metric $\tilde g$.

We can replace this information back into the first field equation and obtain
$$
f' \>\tilde R_{\mu\nu}  -\frac[1/2] f g_{\mu\nu}   = T_{\mu\nu}
\fl{maFinaleq}$$
where $\tilde R_{\mu\nu}$ is now the Ricci tensor of the metric $\tilde g$. 
Let us stress that $f'$ and $f$ are still evaluated along the Ricci scalar $R$ of the original metric $g$.

\CNote{
This equation can be recasted as an Einstein equation for the conformal metric $\tilde g$ 
$$
\tilde G_{\mu\nu}:=\tilde R_{\mu\nu}  - \frac[1/2] \tilde R \>\tilde g_{\mu\nu}= \frac[1/f'] \(T_{\mu\nu}  -\frac[1/2] \(  (f')^2 \tilde R-f \)g_{\mu\nu}  \)
=\frac[1/f'] T_{\mu\nu} + T^{(g)}_{\mu\nu}
\fl{www}$$
with an additional effective source $T_{\mu\nu}^{(g)}= -\frac[1/2f'] \(  (f')^2 \tilde R-f \)g_{\mu\nu}$.
The conservation laws for matter follow from Bianchi identities of $\tilde g$ and read as follows
$$
\tilde \na_\mu\(\frac[1/f'] T^{\mu\nu} + (T^{(g)})^{\mu\nu}\)=0
\fn$$
where $\tilde \na$ is the covariant derivative with respect to $\tilde g$.

Let us stress that the effective energy-momentum tensor $T_{\mu\nu}^{(g)}$, being {\it effective}, does not need to obey other separated 
physical energy conditions which are usually required for matter energy-momentun tensors. 
In this sense one can hope to choose $f$ so that the effective energy-momentum tensor mimick dark energy/matter without needing to introduce extra exotic matter fields. Of course, the freedom in choosing matter fields to model the dark side of the universe is transformed into the freedom in choosing the function $f$.

The standard universality  result has been proved in the metric-affine formulation \ShowLabel{MetricAffineLag}; it holds in vacuum (i.e.~$T_{\mu\nu}=0$ and, more generally, when the trace $T$ vanishes); 
the master equation reads thence as
$$
f' R -2f=0
\fn$$
which, except few particular cases and the degenerate case $f=R^2$ that makes it an identity, has isolated zeros; by choosing $\rho_0$ to be one of the simple zeroes then one must have $R=\rho_0$ on-shell
and the field equation can be recasted as
$$
 f' \(\tilde R_{\mu\nu}  -\frac[1/4]  R g_{\mu\nu} \)  =0
 \quad\then
 \tilde R_{\mu\nu}  -\frac[1/2] \tilde R \tilde g_{\mu\nu} =-\frac[1/4 |f'|] \rho_0 \tilde g_{\mu\nu}  =\La \tilde g_{\mu\nu} 
\fn$$
which are in fact Einstein equations with cosmological constant $\La=-\frac[1/4|f'|] \rho_0$.
Multiple zeroes are discussed in \ref{Universality}.
}

\NewSection{Universality Theorem}

Let us now consider the Lagrangian
$$
L= e f( \beR)  + L_{mat} (e, j^1\phi)
\fl{GHLag}$$
where $e=\det(e^a_\mu)$ denotes the determinant of the frame, $f$ is an analytic function such that the so-called {\it master equation}  $ \beR f'( \beR)-2 f( \beR)=T$ can be solved for $\beR(T)$ as in Section $2$.
Here $\phi$ is a set of matter fields; 
the important issue is the hypothesis that the matter Lagrangian is independent of $\Ga^{ab}_\mu$ (while it could depend on the spin connection 
$\om^{ab}_\mu(e)$ induced by the frame).

The variation of the matter Lagrangian defines the matter energy-momentum tensor
$$
-2e T^a_\mu = \Frac[\de L_{mat}/\de e_a^\mu]
\fn$$

Let us start computing the variation of the Lagrangian \ShowLabel{GHLag} with respect to the frame and the connection:
$$
\eqalign{
\de L=& e\Big(-  f   e^a_\mu \de e_a^\mu+ f'   \Big(2\na_\mu \de \Ga^{ab}_{\nu}  e_a^\mu e_b^\nu   + 2R^{b}{}_{\nu} \de e_b^\nu   +\cr
&+2\be \na_\mu \de \Ga^{ab}_{\nu}   e^{c\mu} e^{d\nu} \ep_{cdab} +2 \be  R^{ab}{}_{\mu\nu}   e^{c\mu} \de e^{d\nu} \ep_{cdab}\Big)  
- 2T^a_\mu \de e_a^\mu\Big)=\cr
=& 2e\( f' R^{a}{}_{\mu}  - \frac[1/2] f   e^a_\mu + \be  R^{cd}{}_{\mu\nu}   e_b^{\nu}  \ep_{cd}\^a\^b - T^a_\mu\)\de e_a^\mu     +\cr
&-2 \na_\mu\(e f'   e_c^\mu e_d^\nu\)  \( \de^c_{[a} \de^d_{b]}+\be    \ep\^c\^d{}_{ab} \)   \de \Ga^{ab}_{\nu}
+2\na_\mu\(e f'   e_c^\mu e_d^\nu  \( \de^c_{[a} \de^d_{b]}+\be    \ep\^c\^d{}_{ab}\)    \de \Ga^{ab}_{\nu}\)\cr
}
\fn$$
As usual the boundary term vanishes because of the boundary conditions $\de\Ga^{ab}_{\mu}=0$ and field equations are
$$
\cases{
& f' R^{a}{}_{\mu}  - \frac[1/2] f   e^a_\mu + \be  R^{cd}{}_{\mu\nu}   e_b^{\nu}  \ep_{cd}\^a\^b =T^a_\mu\cr
&\na_\mu\(e f'   e_c^\mu e_d^\nu\)  \( \de^c_{[a} \de^d_{b]}+\be    \ep\^c\^d{}_{ab} \) =0\cr
}
\fl{mafe}$$

Let us set $\Phi^{cd}_{ab}:=\de^c_{[a} \de^d_{b]}+\be\ep\^c\^d{}_{ab}$; being it skew in both the upper and lower pair it defines an
endomorphisms  $\Phi: \La^2\arr \La^2$ in the space of skew $2$-tensors. It can be proved (e.g.~by explicit computation by Maple)
that it is invertible. One can easily check that its inverse is
$$
\(\Phi^{-1}\)^{ab}_{cd}= (1+4\be^2)\( \de^a_{[c} \de^b_{d]}-\be\ep\^a\^b{}_{cd}\)
\fn$$

\CNote{{ 
Here we are discussing Lorentzian sector. In Euclidean signature the special cases $\be=\pm \frac[1/2]$ must be dealt with separately. They correspond to (anti)selfdual cases in which the map $\Phi^{cd}_{ab}$ is degenerate. Accordingly, the analysis needs extra care though it leads to similar results.}
}

The second field equation is then
$$
\na_\mu\(e f'   e_{[c}^\mu e_{d]}^\nu\)=0
\fn$$

Let us now define a new frame
$$
\tilde e^a_\mu= \sqrt{|f'|} e^a_\mu
\fn$$
We shall systematically denote by a tilde the quantities computed in the new frame.

The second field equation is then
$$
\na_\mu\(\tilde e   \tilde  e_{[c}^\mu \tilde e_{d]}^\nu\)=0
\fl{F2}$$
which is the same equation obtained in the standard frame-affine framework; it implies that $\Ga^{ab}_\mu\equiv \tilde \om^{ab}_\mu$
coincides with the spin connection induced by the frame $\tilde e_a^\mu$.

\CNote{
Equation \ShowLabel{F2} can be recasted as:
$$
\na_\mu\(\tilde e   \tilde  e_c^\mu \tilde e_d^\nu \ep^{abcd}\)=0
\qquad\then
\na_\mu\(  \tilde e^a_\al \tilde e^b_\be \ep^{\al\be\mu\nu} \)=0
\qquad\then
\na_{[\mu}\(  \tilde e^a_\al \tilde e^b_{\be]}  \)=0
\fl{F2b}$$

Let us define $k^{ab}{}\_c= \(\Ga^{ab}_\mu-\tilde \om^{ab}_\mu\) \tilde e^\mu_c$; 
this is the difference between the dynamical connection $\Ga$ and the spin connection induced by the frame $\tilde e$. 
By construction $k^{abc}=- k^{bac}$.
By considering the covariant derivative of the frame we have
$$
\tilde e_c^\mu\na_\mu \tilde e^a_\nu =  \tilde e_c^\mu \(  d_\mu \tilde e^a_\nu -\tilde\Ga^\la_{\nu\mu} \tilde e^a_\la + \Ga^a{}_{b\mu}  \tilde e^b_\nu\)=
\tilde e_c^\mu \(  \tilde\na_\mu \tilde e^a_\nu \)  +k\^a{}_{bc}  \tilde e^b_\nu= k\^a{}_{bc}  \tilde e^b_\nu
\fn$$

Hence equation \ShowLabel{F2b} can be recasted as
$$
\eqalign{
 & \na_{[\mu}\tilde e^{[a}_\al \tilde e^{b]}_{\be]}  =0
\quad\iff
k^{a}{}_{cd}\tilde e^c_{[\al}   \tilde e^d_{\mu} \tilde e^b_{\be]} - k^{b}{}_{cd}\tilde e^c_{[\al}   \tilde e^d_{\mu} \tilde e^a_{\be]} =0\cr
&\iff
k^{a}{}_{eg}	 	\de^b_{f}
+k^{a}{}_{fe}	 	\de^b_{g}
+k^{a}{}_{gf}		\de^b_{e} 
= k^{b}{}_{eg}	 	\de^a_{f} 
+k^{b}{}_{fe}	 	\de^a_{g} 
+k^{b}{}_{gf}		\de^a_{e} 
\cr
&\then
3k^{a}{}_{eg}	 	
+2k^{a}{}_{ge}	 
= k^{b}{}_{gb}		\de^a_{e} 
\quad\then
2k^{a}{}_{ga}	 
= 4k^{a}{}_{ga}  
\quad\then
k^{a}{}_{ga}=0
\cr
}
\fn$$
and substituting back into the original equation we obtain
$$
\eqalign{
&
3k^{a}{}_{eg}	 	
+2k^{a}{}_{ge}	 
= 0
\quad\then
k^{a}{}_{eg}	 	
= -2k^{a}{}_{(ge)}	 
\quad\then
k^{a}{}_{[eg]}	 	
=0}
\fn$$

Finally, we have the following lemma:

\ms
{\ni\bf Lemma:}
if $k_{abc}=-k_{bac}$ and $k_{abc}= k_{acb}$ then $k_{abc}=0$
\ss

{\ni\bf Proof:}
Let us simply notice that
$$
k_{abc}=-k_{bac}= -k_{bca}= k_{cba}= k_{cab}= -k_{acb}=-k_{abc}
\fn$$
from which the thesis follows.\hfill\QDE

\ms

Hence since $k^{ab}{}_\mu=0$ then $\Ga^{ab}_\mu=\tilde \om^{ab}_\mu$.
}

This piece of information can be used to express the curvature tensor in terms of the Riemann tensor of the metric $\tilde g_{\mu\nu}$ induced by the frame 
$\tilde e_a^\mu$:
$$
\eqalign{
&R^{ab}{}_{\mu\nu} = \tilde R^{\al}{}_{\la\mu\nu} \tilde g^{\la\be} \tilde e^a_\al \tilde e^b_\be\cr
&R^{a}{}_{\mu} = R^{ab}{}_{\mu\nu}  e_b^\nu= \sqrt{|f'|} \tilde R^{\al}{}_{\la\mu\nu} \tilde g^{\la\be}\tilde e^a_\al \tilde e^b_\be \tilde e_b^\nu
= \sqrt{|f'|} \tilde g^{\al\be} \tilde R_{\be\mu} \tilde e^a_\al 
=  \ep f' \tilde g^{\al\be} \tilde R_{\be\mu}  e^a_\al
=   \tilde R\^\al{}_{\mu}  e^a_\al\cr
&R=R^{a}{}_{\mu}  e_a^\mu= \ep f' \tilde R
\cr
} 
\fn$$
(Let us stress that in our notation Greek indices are moved by the metric $g$ and not by $\tilde g$; hence $\tilde R\^\al{}_{\mu}$ means $\tilde R\^\al{}_{\mu}= g^{\al\be} \tilde R_{\be\mu}$, not $\tilde g^{\al\be} \tilde R_{\be\mu}$ as usual. This is a consequence of having two metrics around. One should specify which metric is involved at any step!)

Before being ready to manipulate the first field equation we have to prove that the extra term in equation \ShowLabel{mafe} vanishes:
$$
\eqalign{
  R^{cd}{}_{\mu\nu}   e_b^{\nu}  \ep_{cd}\^a\^b
  =&  \sqrt{|f'|}  \tilde R^{\al}{}_{\la\mu\nu} \tilde g^{\la\be}\tilde e^c_{\al} \tilde e^d_{\be}  \tilde e_b^{\nu}  \ep_{cd}\^a\^b
 =  \sqrt{|f'|}  \tilde R^{\al}{}_{\la\mu\nu} \tilde g^{\la\be}\tilde e^c_{\al} \tilde e^d_{\be}  \tilde e^b_\si \tilde g^{\si\nu}  \ep_{cdeb} \eta^{ae}= \cr
 = & \sqrt{|f'|}  \tilde e \tilde R^{\al}{}_{\la\mu\nu} \tilde e^a_\ga   \ep_{\al\be\rho\si} \tilde g^{\la\be}  \tilde g^{\ga\rho}\tilde g^{\si\nu}
 = - \sqrt{|f'|}  (\tilde e)^{-1} \tilde R^{\al}{}_{\la\mu\nu} \tilde e^a_\ga    \tilde g_{\al \be}  \ep^{\be\la\ga\nu} = \cr
 =& - \sqrt{|f'|}  (\tilde e)^{-1} \tilde g_{\mu\al}  \tilde e^a_\ga \>\tilde R^\al{}_{[\nu\be\la]}     \ep^{\be\la\ga\nu} 
=0
}
\fn$$
and that the scalar $\beR$ coindices with the Ricci scalar $R$ of the metric $g$ induced by the original frame $e_a^\mu$:
$$
\eqalign{
  \beR=&  R  +\be R^{ab}{}_{\mu\nu}   e^{c\mu} e^{d\nu} \ep_{cdab} =
  R  +\ep \be  f' \tilde R^{\al}{}_{\la\mu\nu} \tilde g^{\la\be}\tilde e^a_\al \tilde e^b_\be   \tilde e^{c}_\rho \tilde e^{d}_\si \ep_{cdab} \tilde g^{\mu\rho}\tilde g^{\nu\si}=\cr
=&    R  +\ep \be \tilde e  f' \tilde R^{\al}{}_{\la\mu\nu} \ep_{\rho\si\al\be} \tilde g^{\mu\rho}\tilde g^{\nu\si}\tilde g^{\la\be}
= R  -\ep \be (\tilde e)^{-1}  f' \tilde R^{\al}{}_{[\la\mu\nu]}   \tilde g_{\al \be} \ep^{\mu\nu\be\la}=R
\cr
}
\fn$$

The first field equation in \ShowLabel{mafe} can be now recasted as follows
$$
 \(  \ep (f')^2\tilde g^{\al\be} \tilde R_{\be\mu}    - \frac[1/2] f  \de^\al_\mu \)e^a_\al =T^a_\mu
\fn$$
$$
 \(   f' \tilde R\^{\al}{}_{\mu}    - \frac[1/2] f  \de^\al_\mu \)=T^a_\mu e_a^\al =: T\^\al{}_\mu 
\fn$$
$$
f' \tilde R_{\mu\nu}    - \frac[1/2] f   g_{\mu\nu}  =: T_{\mu\nu}
\fn$$
 which coincides with the equation in standard metric-affine $f(R)$-theories (see equation \ShowLabel{maFinaleq}). 
In fact, $f'$ and $f$ are evaluated at the Ricci scalar $R$ of the original metric $g$ and if the matter Lagrangian depends on the frame through its associated metric (as assumed otherwise there is no metric affine formulation to compare with) one has
$$
-2eT^a_\mu = \Frac[\de L_{mat}/\de e_a^\mu]= 2\Frac[\de L_{mat}/\de g_{\mu\nu}] e^{a\nu}= -2\sqrt{g} T_{\mu\nu}e^{a\nu}
\quad\then
T_{\mu\nu}= T^a_\mu e_{a\nu}
\fn$$
We stress that equivalence holds both in vacuum and in the presence of matter.

\NewSection{Conservation Laws}

We shall here compute and discuss conservation laws for $f(\beR)$ theories, following \ref{Libro}. We shall use the formalism introduced in \ref{Augmented}
to which we refer for motivations.
The case of standard Holst action has been already discussed in \ref{HolstCQ}.

The Lagrangian \ShowLabel{GHLag} is gauge-natural (see \ref{Libro}); any generator $\Xi=\xi^\mu\del_\mu + \xi^{ab}\si_{ab}$ of automorphisms on $P$ is 
thence a symmetry.
Here $\si_{ab}$ is a right invariant pointwise basis of vertical vector fields on $P$.
Accordingly, (in vacuum) the following Noether current is conserved
$$
\calE= e\(2 f'   e_c^\mu e_d^\nu  \Phi^{cd}_{ab}  \Lie_{\Xi}\Ga^{ab}_{\nu}-\xi^\mu f\) ds_\mu
\fn$$
where $ds_\mu$ is the local basis of $3$-forms on $M$ induced by coordinates and $ \Lie_{\Xi}\Ga^{ab}_{\mu}= \xi^\nu R^{ab}{}_{\nu\mu}+\na_\mu \xi^{ab}_{(v)} $ denotes the Lie derivative and we set $ \xi^{ab}_{(v)} =\xi^{ab}+ \xi^\mu\Ga^{ab}_\mu$ for the vertical part of the symmetry generator. 
Hence we have
$$
\eqalign{
\calE=& e\(2 f'   e_c^\mu e_d^\nu  \Phi^{cd}_{ab} \( \xi^\la R^{ab}{}_{\la\nu}+\na_\mu \xi^{ab}_{(v)}\)-f\xi^\mu \) ds_\mu=\cr
=&e\( \xi^\la\( 2 f'   e_c^\mu e_d^\nu R^{cd}{}_{\la\nu}
+2\be f'   e_c^\mu e_d^\nu  \ep\^c\^d{}_{ab}R^{ab}{}_{\la\nu}
-\de^\mu_\la f\)+2 f'   e_c^\mu e_d^\nu  \Phi^{cd}_{ab}\na_\mu \xi^{ab}_{(v)}\) ds_\mu=\cr
=&\( 2e\xi^\la\(  f'     R^{c}{}_{\la}-\frac[1/2] f e^c_\la 
+\be f'   e_d^\nu  \ep\^c\^d{}_{ab}R^{ab}{}_{\nu\la}
\)e_c^\mu
-2\na_\mu  \(ef'   e_c^\mu e_d^\nu \) \Phi^{cd}_{ab}\xi^{ab}_{(v)}
\) ds_\mu+\cr
&+\Div\(2 ef'   e_c^\mu e_d^\nu  \Phi^{cd}_{ab} \xi^{ab}_{(v)}ds_{\mu\nu}\)\cr
}
\fn$$
where $ds_{\mu\nu}$ is the local basis of $2$-forms on $M$ induced by coordinates.

Let us define the reduced current $\tilde\calE$ and the superpotential $\calU$
$$
\cases{
&\tilde\calE=\( 2e\xi^\la\(  f'     R^{c}{}_{\la}-\frac[1/2] f e^c_\la 
+\be f'   e_d^\nu  \ep\^c\^d{}_{ab}R^{ab}{}_{\nu\la}
\)e_c^\mu
-2\na_\mu  \(ef'   e_c^\mu e_d^\nu \) \Phi^{cd}_{ab}\xi^{ab}_{(v)}
\) ds_\mu\cr
&\calU=2f'  \>e  e_c^\mu e_d^\nu \> \Phi^{cd}_{ab} \xi^{ab}_{(v)}\>ds_{\mu\nu}
=2\ep\> \tilde e   \tilde e_c^\mu \tilde e_d^\nu\>  \Phi^{cd}_{ab} \xi^{ab}_{(v)}\>ds_{\mu\nu}
\cr
}
\fn$$
Accordingly, the Noether current is 
$$
\calE=\tilde\calE+Div\(\calU\)
\fn$$
Notice that the reduced current vanishes on-shell (see equations \ShowLabel{mafe}). 
Conserved quantities are then generated by integrating the superpotential.

As usual in gauge--natural theories which are equivalent to a natural theory the correspondence among conservation laws is established by
means of the so-called Kosmann lift $\xi^{ab}_{(v)}= \tilde e^a_\al \tilde e^{b\be}\tilde \na_\be \xi^\al$;
see \ref{KL}.
Then the superpotential is:
$$
\eqalign{
\calU=&2\ep \tilde e\(  \tilde e_c^\mu \tilde e_d^\nu  \Phi^{cd}_{ab} \tilde e^a_\al \tilde e^{b\be}\tilde \na_\be \xi^\al \)ds_{\mu\nu}=\cr
=&2\ep \tilde e\(   \tilde \na^\nu \xi^\mu
+\be  \tilde e_c^\mu \tilde e_d^\nu  \ep\^c\^d{}_{ab} \tilde e^a_\al \tilde e^{b\be}\tilde \na_\be \xi^\al \)ds_{\mu\nu}=\cr
=&2 \ep\tilde e\(   \tilde \na^\nu \xi^\mu
+\be  \tilde e_c^\mu \tilde e_d^\nu   \tilde e_a^\al \tilde e_b^{\be}\ep^{cdba} \tilde \na_\be \xi_\al \)ds_{\mu\nu}=\cr
=&2\ep \tilde e  \tilde \na^\nu \xi^\mu \>ds_{\mu\nu}
+2\ep \be \ep^{\mu\nu \be\al} \tilde \na_\be \xi_\al \>ds_{\mu\nu}=\cr
=&2\ep \tilde e  \tilde \na^\nu \xi^\mu \>ds_{\mu\nu}
+\Div  \(\frac[2/3] \ep\be \ep^{\mu\nu \be\al} \xi_\al \>ds_{\mu\nu\be}\)\cr
}
\fn$$
which differs from the Komar superpotential computed for the frame $\tilde e$ by a pure divergence which does not contribute when integrated along closed regions and hence it does not contribute to conserved quantities.

Let us stress that one is forced to choose the Kosmann lift along the new frame $\tilde e$ in order to obtain a correspondence with the conservation laws of
metric-affine formulation.

\NewSection{Conclusions and Perspectives}

The result we obtained can be considered from two points of view. 
From the point of view of classical gravitational theories, it is interesting to have an equivalent formulation of the usual extended theories
$f(R)$ {\it \`a la} Palatini. Using frames allows coupling to spinors; one has just to extend the structure group from $\SO(\eta)$ to $\Spin(\eta)$.
It is also interesting to know that the universality theorem extends further to $f(\beR)$ theories.
Moreover we obtained a non-trivial correspondence among conservation laws. This correspondence selects $\tilde e$ as preferred frame with respect to $e$.
 
From the point of view of Quantum Gravity these models allow a direct approach to quantization {\it \`a la} LQG of all $f(R)$ theories; see \ref{Rovelli}.
Classically these models are known to produce modified dynamics for gravitational physics, in particular in Cosmology.
Even not considering the issue of whether $f(R)$ theories better describe physics than standard GR, it is interesting from the theoretical viewpoint to explore
the dynamics of such an infinite family of models. This could improve the understanding of the emergence of the classical dynamics from the quantum world.
For example it would be interesting to explicitly see whether the formalism developed in these years is able to catch the classical 
difference of dynamics of these models when compared with standard GR. This is particularly relevant in Cosmology where the comparison could help in better  understanding the relation between LQC and LQG; see \ref{Bojowald}.
 
All $f(\beR)$ are gauge--natural theories. As such the gauge and diffeomorphism constraint should be unchanged. 
Accordingly nothing should change in defining the Volume and Area operators together with their quantizations. 
Hence the modified dynamics should appear in Hamiltonian constraint. It will be interesting, at least from a theoretical viewpoint, 
to test the proposals for quantization techniques against therse extended models; see \ref{Thieman}.

A forthcoming paper will be devoded to study the Wheeler-DeWitt equation for the extended models introduced here.

 \

\Acknowledgements

We wish to thank C.~Rovelli for discussions about Barbero-Immirzi formulation.
This work is partially supported by the contribution of INFN (Iniziativa Specifica NA12) 
the local research project {\it Leggi di conservazione in teorie della gravitazione classiche e quantistiche} (2010) 
of Dipartimento di Matematica of University of Torino (Italy).

\ShowBiblio

\end